\documentclass[sigconf, nonacm]{acmart}
\usepackage{amsmath,amsfonts}
\usepackage{graphicx}
\usepackage{textcomp}
\usepackage{xcolor}
\usepackage{algpseudocode}
\usepackage[linesnumbered,lined,ruled,commentsnumbered]{algorithm2e}
\usepackage{xspace}
\usepackage{listings}
\usepackage{color}
\usepackage{multirow}
\usepackage{dblfloatfix}
\usepackage{lscape} 
\usepackage{xparse}
\usepackage{placeins}
\usepackage{caption} 
\usepackage{subcaption}
\usepackage{enumitem}
\usepackage{array, makecell}
\usepackage{listings, xcolor, soul} 
\usepackage{wrapfig}
\usepackage{colortbl}
\usepackage{tabularx} 
\usepackage{booktabs}
\usepackage{courier}
\usepackage{ragged2e}
\usepackage{amsthm}
\usepackage{hyperref}
\usepackage{comment}
\usepackage{pdflscape}
\usepackage{afterpage}
\usepackage{fontawesome}
\usepackage{pifont}
\usepackage{lscape}
\usepackage{enumitem}
\usepackage{tcolorbox}

\setlist[itemize]{noitemsep, topsep=4pt}
\definecolor{lightgray}{HTML}{f6f6f6}
\definecolor{darkgray}{rgb}{.4,.4,.4}
\definecolor{darkblue}{HTML}{800080}
\definecolor{brickred}{HTML}{b04f4f}
\definecolor{purple}{rgb}{0.65, 0.12, 0.82}
\definecolor{diffadd}{HTML}{288f26}
\definecolor{diffrmbg}{HTML}{ffebe9}
\definecolor{diffaddbg}{HTML}{e6ffeb}
\definecolor{diffremove}{HTML}{de4f54}
\definecolor{carrotorange}{rgb}{0.8, 0.33, 0.0}
\definecolor{highlight}{HTML}{fefbc2}
\definecolor{bluegray}{HTML}{3182bd}
\definecolor{delim}{RGB}{20,105,176}
\definecolor{numb}{RGB}{106, 109, 32}
\definecolor{string}{rgb}{0.64,0.08,0.08}

\lstdefinelanguage{JavaScript}{
	keywords={typeof, new, true, false, catch, function, return, null, catch, switch, var, const, let, extends, if, in, while, do, else, case, break, async, await, of},
	keywordstyle=\color{darkblue}\bfseries,
	ndkeywords={class, export, boolean, throw, implements, import, this, setTimeout},
	ndkeywordstyle=\color{brickred}\bfseries,
	identifierstyle=\color{black},
	sensitive=false,
	comment=[l]{//},
	morecomment=[f][\color{diffadd}\bfseries]{+\ },
	morecomment=[s]{/*}{*/},
	morecomment=[f][\color{diffremove}\bfseries]{- },
	commentstyle=\color{violet}\ttfamily,
	stringstyle=\color{carrotorange}\ttfamily,
	morestring=[b]',
	morestring=[b]"
}

\lstdefinelanguage{Python}{
	keywords={typeof, new, true, false, catch, function, return, null, catch, switch, var, const, let, extends, if, in, while, do, else, case, break, async, await, of, from, import, class, def},
	keywordstyle=\color{darkblue}\bfseries,
	ndkeywords={class, export, boolean, throw, implements, import, this, setTimeout, self, __init__},
	ndkeywordstyle=\color{brickred}\bfseries,
	identifierstyle=\color{black},
	sensitive=false,
	comment=[l]{//},
	morecomment=[f][\color{diffadd}\bfseries]{+\ },
	morecomment=[s]{/*}{*/},
	morecomment=[f][\color{diffremove}\bfseries]{- },
	commentstyle=\color{violet}\ttfamily,
	stringstyle=\color{carrotorange}\ttfamily,
	morestring=[b]',
	morestring=[b]"
}

\lstdefinestyle{PythonStyle}{
	language=Python,
	backgroundcolor=\color{lightgray},
	extendedchars=true,
	basicstyle=\scriptsize\ttfamily,
	escapeinside={(*@}{@*)},
	showstringspaces=false,
	showspaces=false,
	numbers=left,
	numberstyle=\scriptsize,
	numbersep=6pt,
	tabsize=4,
	breaklines=true,
	showtabs=false,
	captionpos=b,
	frame=single,
	framesep=4pt,
	linewidth=.98\columnwidth,
	xleftmargin=10pt,
	rulecolor=\color{lightgray}
}

\lstdefinelanguage{json}{
	numbers=left,
	numberstyle=\scriptsize,
	frame=single,
	rulecolor=\color{lightgray},
	backgroundcolor=\color{lightgray},
	showspaces=false,
	showtabs=false,
	breaklines=true,
	postbreak=\raisebox{0ex}[0ex][0ex]{\ensuremath{\color{gray}\hookrightarrow\space}},
	breakatwhitespace=true,
	basicstyle=\ttfamily\scriptsize,
	upquote=true,
	morestring=[b]",
	literate=
	*{0}{{{\color{numb}0}}}{1}
	{1}{{{\color{numb}1}}}{1}
	{2}{{{\color{numb}2}}}{1}
	{3}{{{\color{numb}3}}}{1}
	{4}{{{\color{numb}4}}}{1}
	{5}{{{\color{numb}5}}}{1}
	{6}{{{\color{numb}6}}}{1}
	{7}{{{\color{numb}7}}}{1}
	{8}{{{\color{numb}8}}}{1}
	{9}{{{\color{numb}9}}}{1}
	{\{}{{{\color{delim}{\{}}}}{1}
	{\}}{{{\color{delim}{\}}}}}{1}
	{[}{{{\color{delim}{[}}}}{1}
	{]}{{{\color{delim}{]}}}}{1},
}

\let\origthelstnumber\thelstnumber
\makeatletter
\newcommand*\Suppressnumber{%
	\lst@AddToHook{OnNewLine}{%
		\let\thelstnumber\relax%
		\advance\c@lstnumber-\@ne\relax%
	}%
}

\definecolor{Gray}{gray}{0.6}
\definecolor{LightGray}{gray}{0.9}
\definecolor{bananayellow}{rgb}{1.0, 0.88, 0.5}
\definecolor{myred}{rgb}{1.0,0.44,0.37}

\newcolumntype{a}{>{\columncolor{LightGray}}c}
\newcolumntype{b}{>{\columncolor{LightGray}}r}

\newcommand*\Reactivatenumber[1]{%
	\setcounter{lstnumber}{\numexpr#1-1\relax}
	\lst@AddToHook{OnNewLine}{%
		\let\thelstnumber\origthelstnumber%
	}%
}

\makeatother

\SetCommentSty{mycommfont}

\theoremstyle{definition}

\newcommand{\header}[1]{\par\smallskip\noindent\textbf{#1.}}

\def\BibTeX{{\rm B\kern-.05em{\sc i\kern-.025em b}\kern-.08em
		T\kern-.1667em\lower.7ex\hbox{E}\kern-.125emX}}

\newboolean{showcomments}
\setboolean{showcomments}{true}
\ifthenelse{\boolean{showcomments}}
{
	\definecolor{myyellow}{RGB}{255, 228, 26}
	\definecolor{myblue}{RGB}{50, 50, 220}
	\newcommand{\nb}[2]{
		{\sf
			\fcolorbox{myyellow}{yellow}{\scriptsize\textbf{#1}}%
			$\blacktriangleright$%
			{\color{myblue}\fontsize{7pt}{8pt}\selectfont\textbf{#2}}%
		}%
	}
}
{
	\newcommand{\nb}[2]{}
}

\newcommand{\toolname}{\textsc{Panta}\xspace}

\makeatletter
\DeclareRobustCommand{\change}{%
	\@bsphack
	\leavevmode
	\color{blue}
	\@esphack
}
\DeclareRobustCommand{\stopchange}{%
	\@bsphack
	\normalcolor
	\@esphack
}
\makeatother

\copyrightyear{2026}
\acmYear{2026}
\setcopyright{rightsretained}
\acmConference[ICSE '26]{2026 IEEE/ACM 48th International Conference on Software Engineering}{April 12--18, 2026}{Rio de Janeiro, Brazil}
\acmBooktitle{2026 IEEE/ACM 48th International Conference on Software Engineering (ICSE '26), April 12--18, 2026, Rio de Janeiro, Brazil}
\acmPrice{}
\acmDOI{10.1145/3744916.3764553}
\acmISBN{979-8-4007-2025-3/26/04}

\begin{document}
	\title{LLM Test Generation via Iterative Hybrid Program Analysis
        }
        
        \author{Sijia Gu}
	\affiliation{%
		\institution{University of British Columbia}
		\city{Vancouver}
		\state{BC}
		\country{Canada}}
	\email{sijiagu@ece.ubc.ca}

        \author{Noor Nashid}
	\affiliation{%
		\institution{University of British Columbia}
		\city{Vancouver}
		\state{BC}
		\country{Canada}}
	\email{nashid@ece.ubc.ca}
	
	\author{Ali Mesbah}
	\affiliation{%
		\institution{University of British Columbia}
		\city{Vancouver}
		\state{BC}
		\country{Canada}}
	\email{amesbah@ece.ubc.ca}
    
	\begin{abstract}

Automating unit test generation remains a significant challenge, particularly for complex methods in real-world projects. While Large Language Models (LLMs) have made strides in code generation, they struggle to achieve high branch coverage due to their limited ability to reason about intricate control flow structures. To address this limitation, we introduce \toolname, a technique that emulates the iterative process human developers follow when analyzing code and constructing test cases. \toolname integrates static control flow analysis and dynamic code coverage analysis to systematically guide LLMs in identifying uncovered execution paths and generating better test cases. By incorporating an iterative feedback-driven mechanism, our technique continuously refines test generation based on static and dynamic path coverage insights, ensuring more comprehensive and effective testing. 
Our empirical evaluation, conducted on classes with high cyclomatic complexity from open-source projects, demonstrates that \toolname achieves  26\% higher
line coverage and 23\% higher branch coverage compared to the state-of-the-art.
\end{abstract}
	%
	%

	
	%
	\maketitle

\section{Introduction}
Automated test generation plays a crucial role in software engineering by enhancing code reliability and maintainability. Large Language Models (LLMs) can replicate human coding styles and generate test cases with meaningful variable names, leveraging patterns learned from extensive training data. This capability has led to the adoption of LLMs for test code generation~\cite{cedar:icse23, lemieux:codamosa:icse23, alshahwan2024automated, chen2024chatunitest, pan2025asternaturalmultilanguageunit, symprompt, hits, utgen}.

Despite their effectiveness in general coding tasks, LLMs—trained on diverse datasets—often struggle with the unique code structures and logic of specific projects. This limitation is particularly evident in cases involving complex control flow and conditional statements. Recent studies~\cite{siddiq2024using, schafer2023empirical, elhaji:test-generation-using-copilot:thesis23, wang2024testeval} have shown that LLM-generated test cases frequently fail to achieve sufficient branch coverage, as they inadequately capture intricate conditional logic.

To address this limitation, researchers have explored techniques that supplement LLMs with additional program-related information, such as focal method slices~\cite{hits} and path constraints~\cite{symprompt}. These approaches rely on static prompts containing method fragments to generate tests for each independent segment. However, they lack mechanisms to guide the generation process toward areas requiring additional coverage. Furthermore, since LLMs do not always produce correct code, only a subset of the generated tests compile and pass, leaving significant portions of the code untested. Additionally, these methods primarily focus on generating new test classes rather than augmenting existing ones. CoverUp~\cite{pizzorno2024coverup} takes a different approach by leveraging code coverage information to highlight uncovered portions of the code. However, our empirical evidence suggests that this approach is insufficient for more complex code, where additional contextual guidance is needed to drive meaningful improvements in branch coverage.

	\begin{figure*}[hbt!]
		\centering
	\begin{minipage}[l]{0.34\textwidth}
		\subcaption{Execute existing tests to locate code that lacks coverage}
		\label{fig1-1}
	\begin{lstlisting}
	public static Options parsePattern (String pattern){
		char opt = ' ';(*@\label{line:2}@*)	
		boolean required = false;
		Class<?> type = null;
		final Options options = new Options();
		for (int i = 0; i < pattern.length(); i++){
			final char ch = pattern.charAt(i);
			if (!isValueCode(ch)) {
				// omit some details
				opt = ch;
(*@\bpc@*)			} else if (ch == '!') { (*@\label{line:11}@*)	
(*@\nc@*)				  required = true;(*@\label{line:12}@*)	
			} else {
				type = (Class<?>) getValueClass(ch);
			}
		}
(*@\bpc@*)		if (opt != ' ') {(*@\label{line:17}@*)
			// omit some details
		}
		return options;
	}
	\end{lstlisting}
	\end{minipage}
\hfill
	\begin{minipage}[c]{0.31\textwidth}
		\subcaption{Identify paths and reason about execution scenarios}
		\label{fig1-2}
	\begin{lstlisting}[title={When pattern is empty}, firstnumber=2, backgroundcolor=\color{gray!30}]
	char opt = ' ';
	boolean required = false;
	Class<?> type = null;
	final Options options = new Options();
	for(int i=0;i<pattern.length();i++)(*@{\bf is False}\Reactivatenumber{17}@*)
(*@\bpc@*)	if (opt != ' ') (*@\bf{is False} \Reactivatenumber{20}@*)
	return options;
	\end{lstlisting}
	\begin{lstlisting}[title={When pattern contains only character "!"}, firstnumber=2, backgroundcolor=\color{gray!30}]
	char opt = ' ';
	boolean required = false;
	Class<?> type = null;
	final Options options = new Options();
	for(int i=0;i<pattern.length();i++)(*@{\bf is True}@*)	
	final char ch = pattern.charAt(i);
	if (!isValueCode(ch)) (*@{\bf is False}@*)(*@\Reactivatenumber{11}@*)
(*@\bpc@*)	else if (ch == '!') (*@{\bf is True}@*)	
(*@\nc@*)	required = true;(*@\Reactivatenumber{6}@*)
	for(int i=0;i<pattern.length();i++)(*@{\bf is False} \Reactivatenumber{17}@*)
(*@\bpc@*)	if (opt != ' ') (*@\bf{is False} @*)(*@\Reactivatenumber{20}@*)
	return options;
	\end{lstlisting}
	\end{minipage}
	\hfill
	\begin{minipage}[l]{0.31\textwidth}
	\subcaption{Add new tests for the execution scenarios}
	\label{fig1-3}
	\begin{lstlisting}[numbers=none, framexrightmargin=2pt]
@Test
public void testEmptyPattern() {
	Options options = 
			PatternOptionBuilder.parsePattern("");
	assertTrue(options.getOptions().isEmpty());
}
	
@Test
public void testRequiredOption() {
	Options options = 
			PatternOptionBuilder.parsePattern("!");
	// proper assertions 
}
	\end{lstlisting}
	\end{minipage}
	\caption{Workflow of a developer to add new test cases}
	\label{fig1}
\end{figure*}

In this paper, we present \toolname (\textbf{P}rogram \textbf{AN}alysis guided \textbf{T}est \textbf{A}utomation), a novel approach that combines dynamic code coverage analysis and static control flow analysis to generate tests for under-tested execution paths. Our approach follows an iterative workflow, implementing a multi-step, feedback-driven fully automated process. 
\toolname first gathers dynamic code coverage information and static control flow insights.  
Based on the extracted program analysis information, it identifies and selects execution paths that require additional test coverage for the class, and prompts an LLM to generate test cases that cover those paths. After test generation, \toolname validates the generated tests, adding the passing ones back to the test class while repairing failed tests. The feedback from the test repair process is incorporated into subsequent test generation iterations.

Our work makes the following contributions: 

\begin{itemize}[itemsep=1pt, topsep=2pt, partopsep=0pt]
    \item{An iterative feedback-driven technique that integrates dynamic and static program analysis to assist LLMs in generating tests that cover complex control flow paths.}

    \item{A novel path selection strategy that first ranks linearly independent execution paths—identified via control flow analysis—using a coverage deficiency score computed from code coverage data. It then selects prioritized paths while incorporating historical selection data to effectively balance exploitation and exploration.}

    \item{A fully automated test generation tool, \toolname, that supports both generating new test classes and augmenting existing tests. \toolname is publicly available at \cite{repo}.} 
	
	\item{An empirical evaluation of our approach using real-world open-source programs from the Defects4J~\cite{just2014defects4j} benchmark. Our study demonstrates that \toolname achieves 26\% higher line coverage and 23\% higher branch coverage compared to the state of the art.}
\end{itemize}
	\section{Motivation} 

Recent advances in LLMs have dramatically enhanced their code generation capabilities, including the ability to create tests for diverse execution scenarios. However, when dealing with complex methods that contain multiple conditional branches, LLMs often struggle to achieve comprehensive branch coverage.

As a running example, Figure~\ref{fig1-1}, derived from the Cli project in the Defects4J dataset~\cite{just2014defects4j}, illustrates this challenge. We prompted Llama 3.3 70b~\cite{llama}, Meta’s latest open-source model, to generate tests for the \texttt{parsePattern} method shown in Figure~\ref{fig1-1} and then measured the resulting coverage. The generated tests, however, failed to cover the true branch of the \texttt{else if (ch == ‘!’)} condition at Lines \ref{line:11}--\ref{line:12} and the false branch of the \texttt{if (opt != ’ ’)} condition at Line \ref{line:17}. We also tried with GPT-4o, which similarly missed the false branch of the condition at Line \ref{line:17}. 



Our insight is that emulating a developer’s test creation process can more effectively guide LLMs to generate test cases. Typically, developers begin by checking whether a corresponding test class exists. If it does not, they create a test class skeleton with the necessary imports and gradually add test cases. If it does exist, they run the tests to generate a coverage report that highlights untested statements. They then analyze the control flow—particularly conditional branches and loops—to identify missing execution paths, add new tests to cover these paths, and rerun the tests until adequate coverage is achieved. 

For example, Figure~\ref{fig1-1} shows developers executing the tests and reviewing a JaCoCo~\cite{jacoco} coverage report. In the \texttt{parsePattern} method, the report indicates that two branches remain uncovered. By analyzing the control flow, developers determine that reaching the true branch within a \texttt{for} loop requires the loop condition to be true while accessing the false branch outside the loop depends on the loop condition being false. Figure~\ref{fig1-2} illustrates two representative execution paths: one where an empty pattern prevents loop execution and one where a single \texttt{’!’} triggers the \texttt{else if} branch, which also covers the false branch. New test cases representing these scenarios are shown in Figure~\ref{fig1-3}. 

While developers iteratively refine and expand coverage until achieving the desired adequacy level, manually analyzing the code and creating tests remains a time-intensive task. To address this challenge, our work aims to automate and replicate the iterative workflow that developers follow when generating test cases by combining dynamic and static program analysis to effectively guide LLMs to cover hard-to-cover branches. 

	\begin{figure*}
	\includegraphics[width=0.8\textwidth]{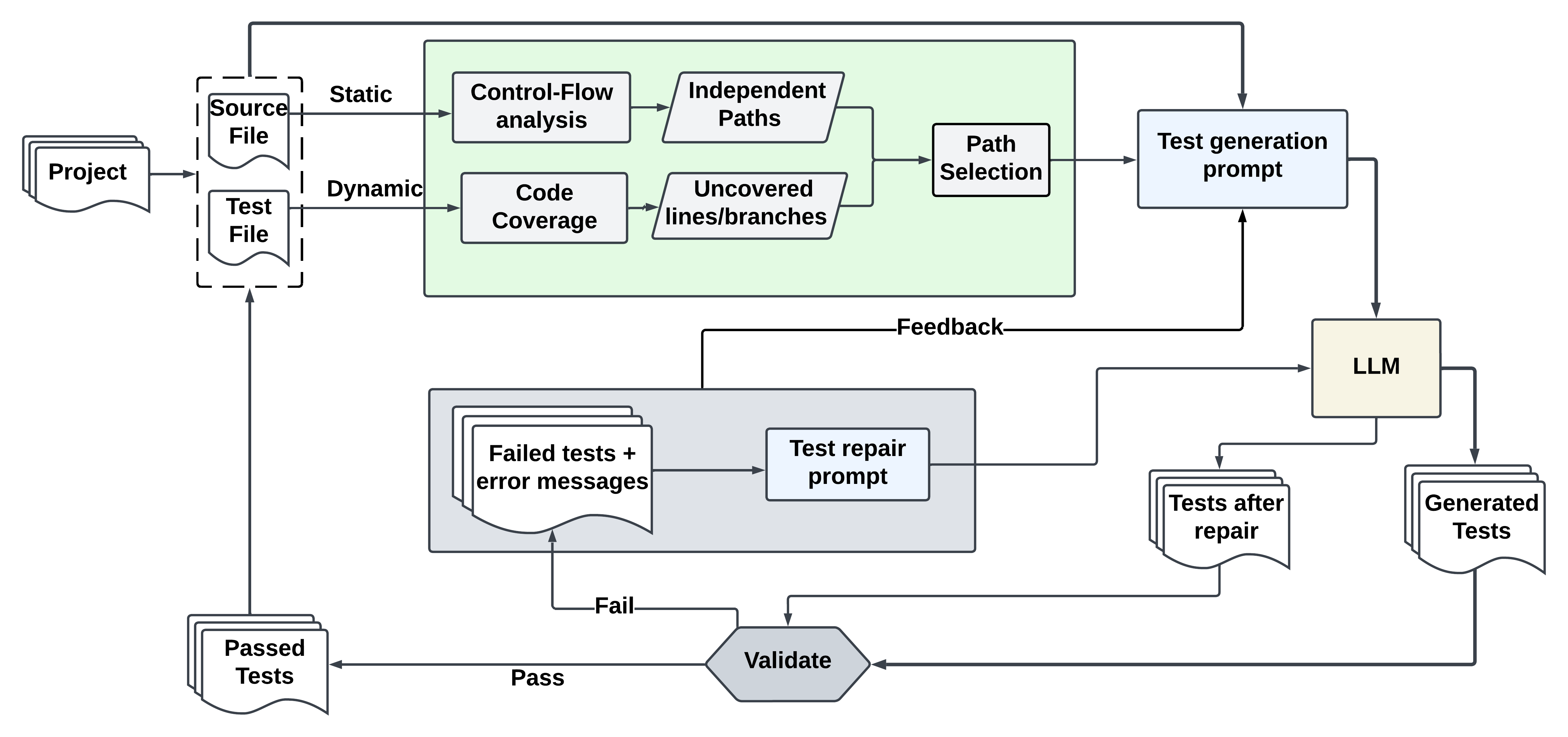}
	\caption{Overview of \toolname.}
	\label{overview}
\end{figure*}

\section{Approach}
Our approach, called \toolname, leverages hybrid program analysis that seamlessly integrates static and dynamic analyses. By iteratively combining static control flow analysis with dynamic code coverage analysis, \toolname enhances LLMs’ ability to determine where and how to generate test cases that improve branch coverage.

Figure \ref{overview} provides an overview of this process. It begins by scanning the entire project to identify directories containing source files and their corresponding test files. If a source file lacks an associated test file, a new test file skeleton is created in the designated test directory. For each source file, the hybrid approach first uses static control flow analysis to extract potential execution paths and then employs dynamic analysis---executing the corresponding test file---to collect real-time branch coverage information. The selected paths, derived from a unique path selection strategy discussed in Section \ref{path-selection}, are incorporated into a test generation prompt that guides the LLM in generating tests.

In addition, the generated tests undergo validation: passing tests are integrated into the test file while failing tests are directed through a test repair procedure. If a test cannot be repaired, it is fed back into the test generation prompt as feedback for the next iteration. This iterative workflow continues until the stopping conditions defined in Section \ref{framework} are met. Since the process is applied uniformly across all source files and their corresponding test files, we detail the iterative test generation process for a single source file and its associated test file in the following subsections.

\begin{algorithm}[t]
	\SetAlgoLined
	\DontPrintSemicolon
	\SetKwFunction{ExplorePaths}{ExplorePaths}
	\SetKwProg{Fn}{Function}{:}{}
	\caption{Path Approximation through CFG}
	\label{alg1}
	\footnotesize{
	\KwIn{srcFile}
	\KwOut{methodDict}	
	$CFG_{f}$ = CFGraph(srcFile)\;\label{alg1_1}
	MUTs = identifyMethodUnderTestFrom(srcFile)\;\label{alg1_2}
	\tcc{extract paths for each method under test}
	\ForEach{method in \text{MUTs}}{
		$CFG_m$ = extractCFGraphForMethod($CFG_{f}$, method)\;\label{alg1_5}
		$CYC_m$= computeCyclomaticComplexity($CFG_m$)\;\label{alg1_6}
		$paths$ = \ExplorePaths{$CFG_m$}\;
		methodDict[method] = ($CYC_m$, $paths$)\;
	}
	\Return methodDict\;
	
	\tcc{Explore paths using BFS}
	 \Fn{\ExplorePaths{$CFG_m$}}{ \label{alg1_12}
	 	start, end = getStartEndNodeFrom($CFG_m$)\;
	 	queue.append([start])\; 
	 	\While{queue is not empty}{
	 		\For{i = 0 to length(queue) - 1}{ 
	 			currentPath = queue.pop(0)\;
	 			lastNode = getLastNodeOf(currentPath)\;
	 			\ForEach{successor in  successors(lastNode)}{
	 				nextPath = copy(currentPath)\;
	 				nextPath.append(successor)\;
	 				\eIf{successor is not the end node}{
	 					queue.append(nextPath)\;
	 				}{
	 					\If{nextPath contains edges not visited}{
	 						methodPaths.append(nextPath)\;
	 					}
	 					\If{all edges of $CFG_m$ has been visited}{
	 						\Return methodPaths\;
	 					}
	 				}
	 			}
	 		}
	 	}\label{alg1_34}
 	}

	}
\end{algorithm}

\subsection{Static Path Approximation}
\label{cfg-prompt}
Extracting all execution paths from a source file is often impractical due to the complexity introduced by loop conditions, which can lead to an exponential or unbounded number of paths. To address this, we propose a path approximation strategy through control flow analysis.

We introduce Algorithm \ref{alg1}, a path approximation method that utilizes the control flow graph (CFG) to extract a set of \emph{linearly independent paths}, ensuring that each path introduces at least one new edge. We leverage an existing CFG generation technique~\cite{comex:arxiv2023} to construct the control flow graph for the entire source file, denoted as $CFG_f$ (Line \ref{alg1_1}). Next, we identify the methods under test (MUTs), excluding helper methods such as private methods, getters, and setters (Line \ref{alg1_2}). For each MUT, we extract its corresponding method-level control flow graph ($CFG_m$) from $CFG_f$ (Line \ref{alg1_5}), forming the basis for our path approximation strategy.


\begin{figure}[h]
	\includegraphics[width=0.4\textwidth]{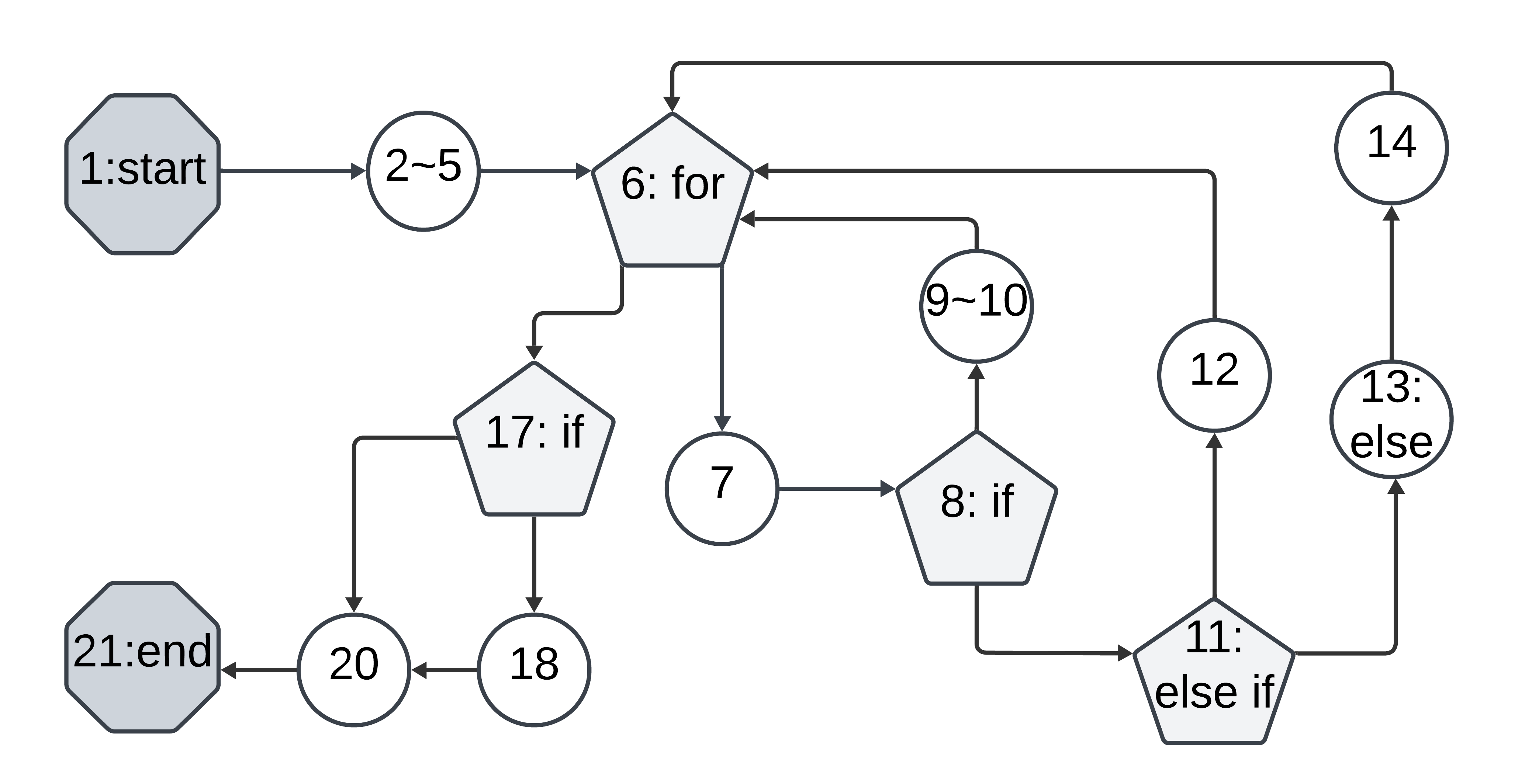}
	\caption{Control flow graph for \texttt{parsePattern} in Figure \ref{fig1}.}
	\label{cfg}
\end{figure}

Each $CFG_m$ is a fully directed graph where code blocks correspond to nodes, and control flow determines the directed edges. We compute the cyclomatic complexity (CYC) \cite{McCabeCyclomatic} for each method (Line \ref{alg1_6}), defined as $\#\text{edges} - \#\text{nodes} + 2$. For instance, in Figure \ref{cfg}, we illustrate the $CFG_m$ for the \texttt{parsePattern} method from Figure \ref{fig1}. We use the line numbers from Figure \ref{fig1} to represent nodes, where lines 2 to 5 and lines 9 and 10 form basic code blocks that are grouped into single nodes. Circles represent nodes without branches, while branch nodes are marked as pentagons. The directed edges indicate the control flow between nodes. To ensure a structured representation, we introduce two additional nodes: a start node (1:start) representing the method entry point and a dummy end node (21:end) to unify all execution paths into a single exit. These start and end nodes are depicted as octagons. Excluding these additional nodes, the \texttt{parsePattern} method consists of 12 nodes and 15 edges, with a cyclomatic complexity (CYC) value of 5. We then extract paths using the \texttt{ExplorePaths} in Algorithm \ref{alg1} (Lines \ref{alg1_12}–\ref{alg1_34}), which is based on breadth-first search (BFS). The algorithm collects the shortest set of paths, ensuring each includes at least one unique edge. Exploration terminates once all edges have been visited. By employing \texttt{ExplorePaths}, we guarantee that for each method under test, the number of extracted paths equals its CYC. 

Using Algorithm \ref{alg1}, we obtain an approximated set of paths for each method under test in a static manner. Next, we use dynamic analysis to identify uncovered execution paths.

\subsection{Dynamic Path Selection}\label{path-selection}

We present a path selection strategy in Algorithm \ref{alg2} that integrates dynamic code coverage data with the static paths extracted from Algorithm \ref{alg1} to identify and prioritize uncovered paths for each test generation iteration.  

Given that multiple test generation iterations may have already occurred, we maintain a \textit{pathHistory} log to track path selection data from previous iterations. Since large language models (LLMs) do not always guarantee the generation of correct tests that fully cover the selected paths in each iteration, a path may be selected multiple times if it remains uncovered. However, if a path has been selected multiple times without being covered, it may indicate that the path itself is invalid or that LLMs struggle to reason about it, making further attempts unproductive. To optimize resource usage, we limit the number of selection attempts per path to \textit{maxSelectedConst}. If a path has been selected \textit{maxSelectedConst} times without being covered, it is excluded from future selections. 
The data variables, \textit{pathHistory} and \textit{maxSelectedConst}, are used globally throughout the entire test generation process. 

\begin{algorithm}[t]
	\SetAlgoLined
	\DontPrintSemicolon
	\SetKwData{pathHistory}{pathHistory}
	\SetKwData{maxSelectedConst}{maxSelectedConst}
	\caption{Path Selection Strategy}
	\label{alg2}
	\footnotesize{
	\KwData{\pathHistory, \maxSelectedConst}
	\KwIn{srcFile, testFile, methodDict}
	\KwOut{selectedPaths}
	
	\tcp{Run tests and collect coverage information}\label{alg2-1}
	runTestsWithCoverage(testFile)\;\label{alg2-2}
	missedLines, missedBranches = parseCoverageReportFor(srcFile)\; \label{alg2-3}	
	selectedPaths = dict()\;
	\tcp{Process each method under test}
	\ForEach{method in methodDict}{\label{alg2-6}	
		methodPaths = methodDict[method][1]\;
		\tcp{Filter paths}
		candidatePaths = []\;
		\ForEach{path in methodPaths}{
			missedScore = count(missedLines $\cap$ path) + count(missedBranches $\cap$ path)\;\label{alg2-11}	
			\If{missedScore > 0}{
				selectedCount = \pathHistory[method][path]\;
				\If{selectedCount < \maxSelectedConst}{
					candidatePaths.append((path, missedScore, selectedCount))\;
				}
			}
		}\label{alg2-18}	
	
		\If{candidatePaths is not empty}{\label{alg2-19}	
			\tcp{Exploitation:  Pick highest missed path}
			highestPath = max(candidatePaths, key=missedScore)\; 
			\pathHistory[method][highestPath] += 1\;
			selectedPaths[method].add(highestPath)\;
			\tcp{Exploration:  Pick least selected path}
			leastPath = min(candidatePaths, key=selectedCount)\; 
			\If{leastPath $\neq$ highestPath}{
				\pathHistory[method][leastPath] +=1\;
				selectedPaths[method].add(leastPath)\;
			}
		}\label{alg2-30}	
	}
	\Return selectedPaths
}
\end{algorithm}

In addition to the global data variables, the input to Algorithm \ref{alg2} consists of the source file, the current test file, and the extracted paths for all methods under test (MUTs) in the source file, denoted as \textit{methodDict}, obtained from Algorithm \ref{alg1}. The output consists of the selected paths for each MUT, which will be incorporated into the prompt for test generation.

After introducing the global, input, and output variables, Algorithm \ref{alg2} begins by executing the current test suite to collect coverage data, parsing the coverage report to identify missed statements and branches in the source file (Lines \ref{alg2-2}–\ref{alg2-3}). For each MUT, we then filter out paths that have been fully covered or have exceeded \textit{maxSelectedConst} selection attempts, retaining the remaining paths as candidates (Lines \ref{alg2-6}–\ref{alg2-18}). These candidate paths have the potential to increase code coverage, as they contain uncovered statements or branches. To prioritize them, we compute a coverage deficiency score for each candidate path: $\text{missedScore} = \text{num of missed statements in path} + \text{num of missed branches in path}$ (Line \ref{alg2-11}). 
We count missed statements as uncovered and partially covered lines; for the two paths in Figure \ref{fig1-2}, there is one missed statement: line 17 for the above path, and three missed statements: lines 11, 12 and 17 for the below path. To enhance the priority of branch coverage, besides the missed statements, we also increment the count of missed branches, which is 1 (above) and 2 (below) for the two paths respectively. Therefore, the \textit{missedScore} for the two paths are 2 (above) and 5 (below), respectively.


For complex methods with high cyclomatic complexity, the large number of candidate execution paths poses a challenge, making it difficult for LLMs to process efficiently and potentially exceed the available context window. For instance, an untested method with a cyclomatic complexity of 20 could have 20 candidate paths, each representing a distinct execution flow. 


To mitigate this, instead of providing all candidate paths in the prompt, we rank them and select two paths per method, based on exploitation and exploration (Lines \ref{alg2-19}--\ref{alg2-30}). For exploitation, we choose the path with the highest \textit{missedScore} to maximize code coverage improvement. For exploration, we select the path that has been visited the least in previous iterations, ensuring diversity in test generation.  If the two selected paths are the same, we include it only once and update \textit{pathHistory} accordingly. This approach ensures that LLMs have a clearer target and can focus more effectively on generating relevant test cases, without exceeding their context window.

\begin{figure}
	\begin{lstlisting}[language=Python, numbers=none, showstringspaces=false]
	## Source File
	Here is the source file that you will be writing tests against.
	{source_file_numbered}
	
	## Test File
	Here is the file that contains the existing tests.
	{test_file}
	
	## Third-party dependencies for test generation
	Please use the following dependencies to generation tests
	{test_dependencies}
	
	{%- if failed_tests_section%}
	{{ failed_tests_section }}
	{% endif %}
	
	## Methods Under Test
	Please generate test for {method_name_1} to cover the path {selected_path_for_method_1}
	...
	Please generate test for {method_name_N} to cover the path {selected_path_for_method_N}
	\end{lstlisting}
	\caption{Code Structure Aware Prompt Template}
	\label{lst:structured-prompt}
\end{figure}

\subsection{Program Analysis Guided Prompt Template} After selecting the paths, we construct a test generation prompt by inserting the selected paths into a structured prompt template. Our program analysis-guided prompt template, shown in Figure~\ref{lst:structured-prompt}, consists of five sections: source file, test file, test dependencies, failed tests, and methods under test.

The source file section includes the complete source code with additional line numbers to facilitate mapping the selected path lines back to their corresponding locations in the code. The test file section contains the currently existing test cases. The test dependencies section specifies the required packages and their versions in the test scope, such as junit:4.13.2, to prevent LLMs from generating tests using incorrect package versions. The failed tests section includes test cases from the last iteration that failed, providing feedback to guide LLMs in refining test generation (\autoref{framework}). Finally, the methods under test section includes the selected paths for each MUT, obtained using Algorithm \ref{alg2}. The paths follow the format illustrated in Figure~\ref{fig1-2}, where each statement is mapped to its corresponding line number in the original source file, and conditional statements are annotated with their branch evaluation (true or false).

\subsection{Iterative Test Generation Framework}\label{framework}

\begin{algorithm}[t]
	\SetAlgoLined
	\DontPrintSemicolon
	\caption{Iterative Test Generation Framework}
	\label{alg3}
	\footnotesize{
		\KwIn{srcFileDir, testFileDir}
		testDeps = extractDependenciesForTestScope()\;\label{alg3-1}
		srcFile = readFile(srcFileDir)\;
		\eIf{testFileDir does not exist}{
			testFile = initialTestClassSkeleton(testFileDir, testDeps)\;
		}{
			testFile = readFile(testFileDir)\;
		}\label{alg3-7}
		iter, iterNoIncrease = 0, 0\;\label{alg3-8}
		pathHistory = dict(), failedTestFeedback = []\;\label{alg3-9}
		methodDict = \textbf{Algorithm~\ref{alg1}} (srcFile)\;\label{alg3-10}
		maxCYC = getMaxComplexity(methodDict)\;\label{alg3-11}
		curCoverage = runTestsWithCoverage(testFile)\;\label{alg3-12}
	
		\While{iter < maxCYC and iterNoIncrease < maxNoIncreaseLimit and curCoverage < 100\% }{\label{alg3-13}
			\tcp{build prompt for test generation}
			\eIf{curCoverage = 0\%}{\label{alg3-15}
				prompt = buildPrompt(srcFile, testFile, testDeps)\;
			}{\label{alg3-17}
				selectedPaths = \textbf{Algorithm~\ref{alg2}} (srcFile, testFile, methodDict, pathHistory)\;\label{alg3-18}
				prompt = buildPrompt(srcFile, testFile, testDeps, selectedPaths, failedTestFeedback)\;
			}\label{alg3-20}
			
			generatedTests = generateTestByPromptLLM(prompt)\;
			
			\tcp{Validate generated tests}
			\ForEach{test in generatedTests}{\label{alg3-23}
				stdout = addNewTestAndRun(test, testFile)\;
				\If{test is failed}{
					errorMsg = parseErrorMessage(stdout)\;
					removeAddedTest(test, testFile)\;
					testsToRepair.add((test, errorMsg))\;
				}
			}\label{alg3-30}
			failedTestFeedback = repairAndValidate (testsToRepair, numOfAttempts)\;\label{alg3-31}
			newCoverage = runTestsWithCoverage(testFile)\;\label{alg3-32}
			\eIf{newCoverage > curCoverage}{
				iterNoIncrease = 0\;
			}{
				iterNoIncrease += 1\;
			}
			curCoverage = newCoverage, iter += 1\;
		}\label{alg3-38}
	}
\end{algorithm}
In the previous subsections, we introduced the algorithms and components used to construct the hybrid program analysis-guided prompt for test generation, focusing on a single iteration.  Now, we elaborate on the design of our iterative test generation workflow, which is presented in Algorithm \ref{alg3} and explain how the various components integrate into the overall \toolname framework.

\header{Test Generation} Algorithm \ref{alg3} begins by preparing the source file, target test file, and test dependencies (Lines \ref{alg3-1}--\ref{alg3-7}). For example, if the project uses Maven, we can extract test dependencies using the command \textit{mvn dependency:list}. We then read the source file as context from the directory $srcFileDir$. Our approach supports two scenarios: (1) augmenting existing tests, and (2) generating a new test file from scratch. We first check whether a test file exists at $testFileDir$. If a test file is available, we analyze and augment the existing test case; otherwise, we create a new test file based on the extracted test dependencies, initializing it with the test class declaration and necessary import statements.

We initialize workflow tracking variables (Lines \ref{alg3-8}--\ref{alg3-9}) to monitor the stopping conditions. Specifically, we use \textit{iter} to count the total number of iterations and \textit{iterNoIncrease} to track the number of consecutive iterations without code coverage improvement. We hypothesize that if several consecutive iterations fail to enhance code coverage, it likely indicates convergence to a fixed value, making further execution unnecessary. Path selection data is stored in \textit{pathHistory}, which is referenced by Algorithm \ref{alg2} for selecting paths. Additionally, we maintain \textit{failedTestFeedback} to store failed tests from the previous iteration, providing feedback to refine the current test generation process. Next, we perform program analysis to extract paths for MUTs using Algorithm \ref{alg1} and obtain the initial code coverage information (Lines \ref{alg3-10}--\ref{alg3-12}). We calculate the maximum cyclomatic complexity among all MUTs in the source file, denoted as \textit{maxCYC} (Line \ref{alg3-11}). Since the number of paths for any MUT is at most \textit{maxCYC}, we use this value to approximate the complexity of the source file and set it as the maximum number of iterations for test generation. This ensures that the iterative workflow runs for at most \textit{maxCYC} iterations. Along with this iteration limit, the process terminates under three conditions (Line \ref{alg3-13}): (1) if the number of iterations reaches the limit \textit{maxCYC}, (2) if the code coverage reaches 100\%, or (3) if there are \textit{maxNoIncreaseLimit} consecutive iterations without coverage improvement. These conditions help optimize resource usage and prevent unnecessary test generation attempts.

For each iteration, we determine the test generation strategy based on the current code coverage. If the coverage is 0\%, indicating that no existing tests are available, we prompt LLMs to generate tests using only the source file, test file, and test dependencies, without providing the selected paths. This approach allows greater flexibility for LLMs to generate tests covering different parts of the source file (Lines \ref{alg3-15}–\ref{alg3-17}). If the coverage is greater than 0\%, we guide LLMs using our program analysis-guided prompt, which includes the selected paths to focus test generation on uncovered parts of the code (Lines \ref{alg3-18}–\ref{alg3-20}). 

\header{Test Validation}
After the LLM completes test generation, the generated tests are directed to the validation step (Lines \ref{alg3-23}--\ref{alg3-30}). The tests are executed one by one. Passed tests are added to the test file, while failed tests, along with their error messages, are collected for the test repair process (Line \ref{alg3-31}). We prompt the LLM with the failed tests and their corresponding error messages, allowing it to attempt repairs for up to \textit{numOfAttempt} times per failed test. We observed that if a test initially fails due to a compilation error, fixing the compilation error may reveal additional compilation errors or runtime errors. Therefore, a single repair attempt is often insufficient. To address this, we set the default value of \textit{numOfAttempt} to 3. The repair process stops when either all failed tests are successfully fixed or the maximum number of repair attempts is reached. After the repair attempts, if any tests still fail, we store them along with their error messages in \textit{failedTestFeedback}, which serves as feedback for the next round of test generation. Finally, we execute the current test file to obtain the updated coverage information and update the \textit{iter} and \textit{iterNoIncrease} variables (Lines \ref{alg3-32}–\ref{alg3-38}). If none of the stop conditions have been met, the process continues with the next iteration.
\section{Evaluation}\label{sec:evaluation}
We assess \toolname  to answer the following three research questions:

\begin{itemize}
    \item \textbf{RQ1:} How does the quality of tests generated by \toolname compare to state-of-the-art test generation techniques?
    
    \item \textbf{RQ2:} What is the individual contribution of each component of \toolname to the overall quality of the generated tests?
    
    \item \textbf{RQ3:} How do different LLMs affect the performance of \toolname in test generation?
\end{itemize}

\begin{table}
	\caption{Included subjects from Defects4J}
	\label{tab:real-sub}
	\small
	\begin{center}
		\setlength\tabcolsep{2pt} 
		\begin{tabular}{l|l|cc|cc} \hline
			{\bf Identifier} & {\bf Project}& {\bf \#Class} & {\bf \#MUTs} & $\textbf{CYC}_{max}$ & $\textbf{CYC}_{med}$\\
			\hline
			Cli&Cli-40f&2&31&11&1\\
			Codec&Codec-18f&7&78&31&1\\
			Collections&Collections-28f&5&218&32&2\\
			Compress&Compress-47f&9&65&22&3\\
			Csv&Csv-16f&3&72&25&1\\
			Gson&Gson-16f&4&75&37&2\\
			JCore&JacksonCore-26f&9&161&30&4\\
			JDatabind&JacksonDatabind-112f&9&371&22&2\\
			JXml&JacksonXml-5f&4&118&29&2\\
			Jsoup&Jsoup-93f&8&171&22&1\\
			JxPath&JxPath-22f&12&115&32&4\\
			Lang&Lang-4f&17&586&30&2\\
			Math&Math-2f&30&452&31&2\\
			Time&Time-13f&11&458&30&2\\
			\hline
			&Total/Ave&130&2971&27&2\\
			\hline
		\end{tabular}
	\end{center}
\end{table}

\subsection{Experimental Setup}\label{setup}
\noindent\textbf{Subject Programs.} We evaluate \toolname using real-world subjects from Defects4J~\cite{just2014defects4j}. Specifically, we selected the latest fixed version of each subject program from Defects4J v2.0.1. To maintain consistency, we exclude three outdated projects---Chart, Mockito, and Closure---as they do not use Maven and rely on deprecated dependencies. The remaining 14 subjects are compatible with Java 8 or higher, Maven (v3.6.3), and JUnit (v4.13.2). They are further integrated with JaCoCo (v0.8.11)~\cite{jacoco} for code coverage measurement and Pitest (v1.17.0)~\cite{coles2016pit} for mutation score evaluation.

To evaluate coverage on complex code, we did not generate tests for all classes; instead, we selected non-abstract, public classes that contain at least one method with cyclomatic complexity greater than 10~\cite{McCabeCyclomatic,sanusi2020development}. Our process began by automatically selecting all public, non-abstract classes from the 14 Defects4J projects that include at least one non-private method exceeding this threshold. We then manually excluded two types of outliers: (1) classes with extreme complexity (e.g., cyclomatic complexity > 40) caused by large switch-statement patterns, and (2) classes with minimal explicit control flow due to heavy reliance on inheritance. These exclusions ensured that our evaluation remained focused on complex yet analyzable control-flow structures. Table~\ref{tab:real-sub} presents the selected subjects. The column Project denotes the class name along with its version number from Defects4J. The column \#Class represents the number of classes for which we generate tests, while \#MUTs counts the total number of methods under test within the selected classes, for each project. In total, we evaluate 130 classes containing 2,971 methods under test (MUTs). Additionally, we report the maximum and median cyclomatic complexity values of the methods under test in the columns $\text{CYC}_{max}$, $\text{CYC}_{med}$, respectively. More details about individual classes are publicly available in our repository~\cite{repo}.

\begin{table*}[h]
	\caption{Comparing \toolname with SymPrompt}
	\label{tab:comparison}
	\small
	\setlength\tabcolsep{3pt}
	\begin{center}
		\begin{tabular}{@{}llaaccaaccbb@{}} \toprule
			\multirow{2}{*}{\bf Identifier}&\multirow{2}{*}{\bf \#Class}&\multicolumn{2}{c}{\bf line coverage (\%)}&\multicolumn{2}{c}{\bf branch coverage (\%)}
			&\multicolumn{2}{c}{\bf pass rate (\%)}&\multicolumn{2}{c}{\bf mutation score (\%)}&\multicolumn{2}{c}{\bf HCC(\#)}\\ \cmidrule(l){3-4}\cmidrule(l){5-6}\cmidrule(l){7-8}\cmidrule(l){9-10}\cmidrule(l){11-12}
			&&\toolname&  SymPrompt &\toolname&  SymPrompt&\toolname&  SymPrompt&\toolname&  SymPrompt&\toolname&  SymPrompt\\ 	
			Cli&2&\textbf{83.56}&42.93&\textbf{70.02}&34.02&\textbf{85.21}&29.31&\textbf{39.31}&33.53&\textbf{1}&0\\
			Codec&7&\textbf{81.83}&47.67&\textbf{77.05}&41.77&\textbf{72.82}&29.04&\textbf{26.10}&19.14&\textbf{5}&1\\
			Collections&5&\textbf{72.66}&61.77&\textbf{68.53}&64.48&40.01&\textbf{49.83}&\textbf{63.43}&55.56&2&\textbf{3}\\
			Compress&9&\textbf{56.95}&21.90&\textbf{46.19}&17.31&\textbf{43.40}&20.29&\textbf{25.16}&9.09&\textbf{1}&0\\
			Csv&3&\textbf{76.99}&46.47&\textbf{56.74}&35.48&\textbf{69.88}&34.45&\textbf{69.45}&48.45&0&0\\
			Gson&4&\textbf{77.07}&40.81&\textbf{60.95}&32.57&\textbf{72.96}&10.44&\textbf{46.17}&23.74&\textbf{1}&0\\
			JCore&9&\textbf{68.43}&36.21&\textbf{60.88}&33.58&\textbf{61.74}&36.32&\textbf{29.79}&12.13&\textbf{3}&1\\
			JDatabind&9&\textbf{56.77}&35.64&\textbf{47.10}&29.93&\textbf{28.99}&23.11&35.34&\textbf{39.85}&\textbf{1}&0\\
			JXml&4&\textbf{67.81}&26.32&\textbf{62.99}&22.26&\textbf{64.18}&26.31&\textbf{41.10}&7.82&1&1\\
			Jsoup&8&\textbf{82.99}&36.99&\textbf{61.98}&25.62&\textbf{68.93}&22.92&\textbf{56.69}&32.89&\textbf{2}&0\\
			JxPath&12&\textbf{56.80}&24.40&\textbf{49.05}&22.48&\textbf{50.97}&14.08&\textbf{45.97}&20.73&\textbf{4}&1\\
			Lang&17&\textbf{71.18}&57.71&\textbf{64.13}&52.18&\textbf{72.27}&51.39&39.01&\textbf{53.31}&\textbf{5}&2\\
			Math&30&\textbf{70.94}&50.45&\textbf{62.81}&44.75&\textbf{41.38}&35.57&\textbf{43.05}&36.68&\textbf{13}&6\\
			Time&11&\textbf{80.57}&58.66&\textbf{70.34}&47.41&\textbf{71.86}&45.05&\textbf{52.45}&45.33&\textbf{6}&1\\
			\hline
			Total/Ave&130&\textbf{70.18}&43.91&\textbf{60.83}&38.17&\textbf{55.92}&32.83&\textbf{43.79}&31.30&\textbf{45}&16\\
			\bottomrule
		\end{tabular}
	\end{center}
\end{table*}

\noindent\textbf{Evaluation Metrics.} We adopt commonly used quality metrics in test generation literature~\cite{hits, utgen, symprompt}, including \textit{line coverage}, \textit{branch coverage}, and \textit{pass rate} for RQ1 and RQ2. The pass rate is defined as the percentage of generated tests that successfully pass when added to the test suite. Since we evaluate at the class level rather than the method level, we measure line and branch coverage per class and report the average coverage across all targeted classes for each project. For RQ1 and RQ3, in addition to these standard metrics, we use two additional metrics: \textit{High Coverage Count (HCC)} and \textit{mutation score}. Based on industry standards~\cite{geekforgeeks-coverage, atlassian-coverage}, typical code coverage targets range from 70\% to 80\%, e.g., Salesforce sets a 75\% threshold for line coverage~\cite{salesforce-coverage}. However, since line coverage is generally easier to achieve than branch coverage, we focus primarily on branch coverage. Therefore, we report the number of classes that achieve at least 75\% branch coverage as a high coverage count, denoted as HCC. Additionally, we evaluate \textit{mutation score} to assess test effectiveness, as coverage alone does not ensure fault detection if test assertions are non-existent or weak~\cite{yucheng:fse15}.

\noindent\textbf{LLM Selection.} We select the latest open-source model, Meta Llama 3.3 70B~\cite{llama}, as our primary model for evaluation. For RQ3, to assess the impact of different models, we evaluate \toolname on three additional models: one open-source model, Mistral Large 2~\cite{mistral} from Mistral AI, and two proprietary cost-efficient models—GPT-4o Mini from OpenAI~\cite{gpt-mini} and Claude 3.5 Haiku from Anthropic~\cite{claude}. All LLMs evaluated support a context window of 128k tokens. The model parameters are held constant across all four LLMs used in our evaluation, with max\_tokens set to 4096 for output generation and temperature set to 0.2.

\toolname supports both augmenting existing test classes and generating new ones. To ensure a fair comparison, we evaluate \toolname in the scenario where tests are generated for a new class without any existing test cases across all research questions. Given the low temperature, the generation process is largely deterministic, with only minor fluctuations. As a result, we perform a single run per class and report the average results across all classes for each project. 

\begin{table*}
	\caption{Ablation Study}
	\label{tab:ablation}
	\small
	\setlength\tabcolsep{1.5pt}
	\begin{center}
		\begin{tabular}{@{}llaaaaccccaaaa@{}} \toprule
			\multirow{2}{*}{\bf Identifier}&\multirow{2}{*}{\bf \#Class} &\multicolumn{4}{c}{\bf line coverage (\%)}&\multicolumn{4}{c}{\bf branch coverage (\%)}
			&\multicolumn{4}{c}{\bf pass rate (\%)}\\ \cmidrule(l){3-6}\cmidrule(l){7-10}\cmidrule(l){11-14}
			&&baseline& {\toolname}\_basic& {\toolname}\_cov & \toolname&  baseline& {\toolname}\_basic& {\toolname}\_cov & \toolname&baseline& {\toolname}\_basic& {\toolname}\_cov & \toolname\\ 	
			Cli&2&24.29 &78.72&\textbf{85.12}&83.56 &13.12&63.99&\textbf{71.93}&70.02&28.57&77.64&66.25&\textbf{85.21}\\
			Codec&7&39.06&77.91&65.55&\textbf{81.83}&32.24&69.46&57.86&\textbf{77.05}&41.51&62.67&59.85&\textbf{72.82}\\
			Collections&5&30.30&66.85&64.97&\textbf{72.66}&29.90&63.74&66.32&\textbf{68.53}&60.80&\textbf{76.38}&75.04&40.01\\
			Compress&9&17.76&47.40&47.59&\textbf{56.95}&14.02&39.06&39.86&\textbf{46.19}&19.88&\textbf{49.90}&40.54&43.40\\
			Csv&3&12.11&66.18&59.34&\textbf{76.99}&6.00&51.25&44.49&\textbf{56.74}&28.70&61.11&61.02&\textbf{69.88}\\
			Gson&4&28.00&62.00&75.46&\textbf{77.07}&17.00&52.30&55.85&\textbf{60.95}&52.92&64.47&64.46&\textbf{72.96}\\
			JCore&9&25.47&40.70&42.37&\textbf{68.43}&15.89&33.17&35.02&\textbf{60.88}&42.60&42.65&45.32&\textbf{61.74}\\
			JDatabind&9&17.90&40.49&38.43&\textbf{56.77}&12.17&30.95&29.74&\textbf{47.10}&37.10&\textbf{47.37}&43.23&28.99\\
			JXml&4&12.26&53.70&38.08&\textbf{67.81}&4.04&40.89&30.37&\textbf{62.99}&25.22&60.71&37.50&\textbf{64.18}\\
			Jsoup&8&42.35&65.80&77.62&\textbf{82.99}&27.99&49.85&61.02&\textbf{61.98}&55.90&52.41&65.19&\textbf{68.93}\\
			JxPath&12&30.77&53.28&52.35&\textbf{56.80}&23.89&45.41&45.42&\textbf{49.05}&42.08&\textbf{59.60}&47.51&50.97\\
			Lang&17&34.47&60.90&58.72&\textbf{71.18}&29.85&53.74&51.34&\textbf{64.13}&65.31&67.29&64.92&\textbf{72.27}\\
			Math&30&37.47&66.73&62.35&\textbf{70.94}&29.39&56.14&55.58&\textbf{62.81}&\textbf{61.84}&55.69&54.96&41.38\\
			Time&11&55.47&71.42&68.84&\textbf{80.57}&40.96&59.40&57.84&\textbf{70.34}&\textbf{86.71}&65.77&66.53&71.86\\
			\hline\hline
			$\text{CYC}{max} \leq 20$ &104&34.01&61.56&60.07&\textbf{70.37}&26.29&51.80&51.86&\textbf{61.09}&51.42&\textbf{58.21}&55.60&56.10\\
			$\text{CYC}_{max} > 20$&26&27.38&55.49&52.68&\textbf{69.42}&19.17&46.00&42.93&\textbf{59.81}&56.55&\textbf{59.15}&56.66&55.16\\
			\hline
			Total/Ave&130&32.69&60.34&58.59&\textbf{70.18}&24.87&50.64&50.08&\textbf{60.83}&52.45&\textbf{58.40}&55.81&55.92\\
			\bottomrule
		\end{tabular}
	\end{center}
\end{table*}

\subsection{RQ1: Generated Test Quality Comparison}
Search-based tools such as EvoSuite~\cite{evosuite:fse11} have already been extensively studied in comparison with LLM-based test generation~\cite{hits,chen2024chatunitest,symprompt,rao2023cat,utgen}. Prior results demonstrate that LLM-based tools have advanced beyond search-based approaches by achieving higher average line and branch coverage, although EvoSuite remains competitive on certain projects. Moreover, search-based generated tests are often difficult to interpret and lack meaningful assertions compared to those produced by LLM-based methods~\cite{rao2023cat,utgen}. Because our contribution lies in advancing LLM-based test generation, we therefore focus our comparisons on other LLM-based tools. The latest test generation techniques relevant to \toolname include HITS~\cite{hits} and SymPrompt~\cite{symprompt}. HITS decomposes focal methods into slices and generates tests slice by slice, while SymPrompt extracts path constraints as a list of branch conditions for each method under test and generates tests path by path. Neither approach leverages dynamic code coverage data and both operate statically without guidance on where additional tests are needed. Consequently, HITS and SymPrompt are limited to generating new test classes rather than augmenting existing ones. To differentiate \toolname with them, \toolname adopts an iterative workflow, guiding the LLM to generate tests for selected paths based on both dynamic and static program analysis information. Our approach works for both scenarios to augment existing tests and add new test classes. 

However, neither HITS nor SymPrompt has publicly available implementations. HITS relies on an LLM to generate slices instead of traditional program analysis, and the paper~\cite{hits} provides limited implementation details, making replication challenging. In contrast, SymPrompt 
provides sufficient details to replicate its approach. Therefore, to compare it with \toolname, we implemented SymPrompt's technique following the information that is available in their paper~\cite{symprompt}.

We present the results in Table~\ref{tab:comparison}. We highlight the best values in bold to enhance visualization in the result tables. In terms of line and branch coverage, \toolname consistently outperforms SymPrompt across all projects. 

On average, \toolname achieves 26.3\% higher line coverage and 22.7\% higher branch coverage than SymPrompt. Regarding high coverage count (HCC), 45 out of 130 classes achieve branch coverage above 75\%, compared to only 16 for SymPrompt. The pass rate of SymPrompt is, on average, 23.1\% lower than \toolname. Beyond the absence of a test repair phase, this can be attributed to its limited prompt context. Specifically, SymPrompt provides only branch conditions rather than the full execution path, making it more challenging for the LLM to infer correct test inputs. In terms of mutation score, \toolname achieves a higher average score (43.8\%) than SymPrompt (31.3\%), aligning with its superior code coverage results.

There are a few scenarios in which SymPrompt performs better. In the Collections project, SymPrompt achieves a higher pass rate than \toolname, while in JDatabind and Lang, its mutation score exceeds that of \toolname. Upon analysis, these scenarios occur in large utility classes containing large number of methods. For example, \texttt{MapUtils} in Collections has 71 methods. Since \toolname generates tests at the class level, it selects paths for each MUT that have not been fully covered in each iteration. For large classes, this results in long prompts, increasing the difficulty for the LLM to generate complete tests. In contrast, SymPrompt operates at the method level, allowing it to focus on smaller contexts, making it more effective in these specific cases. 

Overall, across all evaluation metrics---line coverage, branch coverage, pass rate, HCC, and mutation score---\toolname consistently outperforms SymPrompt.

\subsection{RQ2: Ablation study}
To assess the contributions of each component of \toolname, we design different variants that provide the LLM with varying levels of information for test generation. We evaluate \toolname in a scenario where tests are generated for a new class without any existing test cases. As shown in Algorithm~\ref{alg3}, the essential inputs include the source file (srcFile), test dependencies (testDeps), and a new test file (testFile), initialized with the test class declaration and necessary import statements.

\begin{table*}
	\caption{Comparison among different models}
	\label{tab:models}
	\small
	\setlength\tabcolsep{1.5pt}
	\begin{center}
		\begin{tabular}{@{}laaaaccccaaaaccccbbbb@{}} \toprule
			\multirow{2}{*}{\bf Identifier}&\multicolumn{4}{c}{\bf line coverage (\%)}&\multicolumn{4}{c}{\bf branch coverage (\%)}
			&\multicolumn{4}{c}{\bf pass rate (\%)}&\multicolumn{4}{c}{\bf mutation score (\%)}&\multicolumn{4}{c}{\bf HCC(\#)}\\ \cmidrule(l){2-5}\cmidrule(l){6-9}\cmidrule(l){10-13}\cmidrule(l){14-17}\cmidrule(l){18-21}
			&Llama& Mistral & GPT& Claude&Llama& Mistral & GPT& Claude&Llama& Mistral & GPT& Claude&Llama& Mistral & GPT& Claude&Llama& Mistral & GPT& Claude\\ 	
			Cli&83.56&58.04&75.79&\textbf{97.61}&70.02&39.38&54.41&\textbf{82.33}&85.21&33.52&66.73&\textbf{86.30}&39.31&33.53&47.98&\textbf{74.57}&1&0&0&\textbf{2}\\
			Codec&81.83&73.20&55.24&\textbf{86.20}&77.05&68.00&49.45&\textbf{80.82}&\textbf{72.82}&41.40&43.72&69.48&26.10&36.37&25.35&\textbf{48.14}&\textbf{5}&4&1&\textbf{5}\\
			Collections&72.66&\textbf{78.64}&60.73&61.10&68.53&\textbf{77.61}&58.13&62.95&40.01&51.59&53.64&\textbf{82.73}&63.43&\textbf{70.89}&62.87&62.87&2&\textbf{4}&2&3\\
			Compress&56.95&31.14&37.52&\textbf{74.83}&46.19&27.44&29.13&\textbf{70.08}&43.40&16.99&23.67&\textbf{61.61}&25.16&11.82&13.98&\textbf{30.05}&1&0&0&\textbf{4}\\
			Csv&\textbf{76.99}&69.17&46.39&62.08&56.74&\textbf{58.61}&28.77&46.89&\textbf{69.88}&33.16&24.23&55.13&\textbf{69.45}&67.78&52.98&68.02&0&0&0&0\\
			Gson&77.07&48.32&73.32&\textbf{79.33}&60.95&37.67&56.48&\textbf{61.24}&\textbf{72.96}&20.02&34.58&62.03&46.17&25.29&47.99&\textbf{53.05}&1&1&0&1\\
			JCore&\textbf{68.43}&45.73&44.75&65.90&60.88&40.25&39.27&\textbf{61.04}&61.74&31.65&35.72&\textbf{69.16}&\textbf{29.79}&13.32&13.61&25.48&3&2&1&\textbf{4}\\
			JDatabind&\textbf{56.77}&45.96&52.50&52.45&47.10&42.99&46.70&\textbf{49.70}&28.99&23.22&49.83&\textbf{62.81}&35.34&37.14&35.83&\textbf{38.12}&1&\textbf{2}&1&\textbf{2}\\
			JXml&\textbf{67.81}&41.88&34.59&34.02&\textbf{62.99}&38.66&32.41&38.17&\textbf{64.18}&31.23&31.88&25.11&\textbf{41.10}&17.64&13.04&15.95&1&1&1&1\\
			Jsoup&\textbf{82.99}&60.00&68.25&80.22&61.98&45.75&52.87&\textbf{65.62}&68.93&38.82&48.68&\textbf{75.29}&56.69&42.70&45.45&\textbf{59.65}&2&1&1&\textbf{3}\\
			JxPath&56.80&50.85&32.66&\textbf{65.85}&49.05&42.17&29.20&\textbf{57.78}&50.97&29.07&29.04&\textbf{52.02}&45.97&43.38&29.90&\textbf{50.48}&\textbf{4}&1&1&\textbf{4}\\
			Lang&71.18&65.30&65.00&\textbf{75.61}&64.13&58.85&58.64&\textbf{71.57}&72.27&39.63&56.24&\textbf{82.44}&39.01&32.43&33.05&\textbf{39.42}&5&5&4&\textbf{9}\\
			Math&70.94&56.04&55.02&\textbf{81.27}&62.81&49.62&49.65&\textbf{72.87}&41.38&40.77&51.01&\textbf{73.03}&43.05&28.30&30.56&\textbf{44.86}&13&8&8&\textbf{16}\\
			Time&\textbf{80.57}&70.99&64.26&79.81&70.34&63.61&54.25&\textbf{74.15}&\textbf{71.86}&45.37&47.05&64.81&52.45&51.85&35.46&\textbf{60.24}&6&5&2&\textbf{8}\\
			\hline
			Total/Ave&70.18&56.60&54.05&\textbf{73.20}&60.83&49.85&46.77&\textbf{66.52}&55.92&35.48&44.39&\textbf{68.16}&43.79&36.60&34.86&\textbf{47.92}&45&34&22&\textbf{62}\\
			\bottomrule
		\end{tabular}
	\end{center}
\end{table*}

To establish a baseline, we prompt the LLM with only this basic information for a single test generation pass, without leveraging multiple iterations, test repair, or feedback, which are key features of our iterative framework. To assess the impact of our iterative framework, we introduce {\toolname}\_basic, a variant that generates prompts consistently across all iterations using only the basic information, without incorporating any program analysis data. Additionally, we evaluate {\toolname}\_cov to determine whether code coverage information alone can enhance test coverage. The {\toolname}\_cov variant extends the basic information by incorporating code coverage data, providing the LLM with the locations of uncovered lines and branches in the current class as line numbers mapped to the original source. However, it does not include path selection details on how to cover these uncovered elements.

The results comparing the baseline, {\toolname}\_basic, {\toolname}\_cov, and \toolname are presented in Table~\ref{tab:ablation}. In addition to reporting the average line/branch coverage and pass rate for each project, we classify the 130 evaluated classes based on their maximum cyclomatic complexity ($\text{CYC}_{max}$), using two groups: $\text{CYC}{max} \leq 20$ and $\text{CYC}_{max} > 20$. A cyclomatic complexity greater than 20 is considered an indicator of highly complex programs~\cite{sanusi2020development}.

For line and branch coverage, we observe a substantial improvement when comparing {\toolname}\_basic to the baseline, with line coverage increasing by 27.65\% and branch coverage by 25.77\%. This demonstrates that our iterative framework significantly enhances overall code coverage. While {\toolname}\_cov achieves similar coverage, its average coverage is slightly lower than that of {\toolname}\_basic. A Paired T-test~\cite{marascuilo1988statistical} confirms that the difference between them is not statistically significant across all subjects. Notably, for the Cli project, which has lower complexity ($\text{CYC}_{max} = 11$), {\toolname}\_cov achieves the highest line and branch coverage. This suggests that for less complex subjects, code coverage information alone can be beneficial, but for more complex projects, it is insufficient to drive meaningful improvements in coverage.

By integrating code coverage-driven path selection (Algorithm \ref{alg2}), \toolname further improves line and branch coverage by approximately 10\% compared to the other two \toolname variants. When analyzing coverage across complexity categories, we observe a notable drop of 6\% to 9\% for highly complex subjects ($\text{CYC}_{max} > 20$) in the baseline, {\toolname}\_basic, and {\toolname}\_cov variants. In contrast, \toolname maintains nearly consistent coverage across both complexity levels, with only a 1 percentage point difference. These findings demonstrate that our iterative framework, combined with hybrid program analysis leveraging both static and dynamic information, effectively enhances code coverage, particularly for complex subjects.

The pass rate is primarily determined by the LLM's ability to generate correct code. Since all variants use the same LLM (Llama 3.3 70B), the pass rate remains relatively stable. {\toolname}\_basic achieves a 6\% higher pass rate than the baseline, attributed to the test repair phase incorporated in our framework. However, after integrating program analysis information into the prompt, the overall pass rate is slightly lower than {\toolname}\_basic, with only a 3.5\% improvement over the baseline. This trend holds for both {\toolname}\_cov and \toolname. The reason is that prompts in {\toolname}\_cov and \toolname guide the LLM to target more complex and hard-to-cover execution paths, making it more challenging to generate correct test cases. In contrast, {\toolname}\_basic offers greater flexibility, allowing the LLM to decide where to add tests. As a result, the LLM tends to avoid complex paths, leading to a higher pass rate by focusing on easier-to-cover cases.

\subsection{RQ3: Impact of different models}
In addition to Llama 3.3 70B, we evaluate \toolname on three other models, as introduced in Section~\ref{setup}. Table~\ref{tab:models} presents the results across four models: Llama 3.3 70B (Llama), Mistral Large 2 (Mistral), GPT-4o Mini (GPT), and Claude 3.5 Haiku (Claude) for each evaluation metric.

For code coverage metrics, Claude performs the best, achieving 73.2\% line coverage, 66.5\% branch coverage, and 62 out of 130 classes surpassing 75\% branch coverage. Llama follows, with 70.2\% line coverage, 60.8\% branch coverage, and 45 classes exceeding the 75\% threshold. GPT performs the worst in achieving high coverage. Thus, the ranking for code coverage is Claude > Llama > Mistral > GPT. The same ranking applies to mutation score, as Claude also achieves the highest score, followed by Llama, Mistral, and GPT. For pass rate, which reflects the model’s ability to generate correct code, Claude again performs the best (68.2\%), with Llama as the second highest (55.9\%). However, unlike the coverage metrics, GPT achieves a higher pass rate than Mistral, resulting in a ranking of Claude > Llama > GPT > Mistral for pass rate.

Overall, Claude is the best-performing model, but as a proprietary model, it has the highest API pricing among the four models. For a more cost-effective alternative, Llama achieves results comparable to Claude while being open-source, making it a more budget-friendly option.

\section{Discussion}
\header{Limitations} Regarding the sufficiency of source code context, \toolname analyzes one class at a time, capturing constructors and internal methods but omitting external dependencies across classes. This design is less effective for classes that rely heavily on inheritance, where key behaviors are defined through method overrides in parent classes. In such cases, the absence of explicit method implementations reduces the effectiveness of control flow analysis. However, since our study examines whether hybrid program analysis improves coverage for complex methods and all evaluated prompting strategies operate in a class-level context, the RQ2 ablation study - conducted under identical conditions - demonstrates a clear improvement with hybrid analysis, validating its effectiveness even within this limited scope. In the future, we plan to extend our approach to incorporate cross-class context to further address this limitation. Regarding test repair, a common limitation across different LLMs is their inability to effectively fix runtime errors, such as \textit{AssertionError} caused by failed test assertions. In contrast, LLMs perform better in resolving compilation errors. This indicates that LLMs lack the capability to fully understand project-specific execution logic, as they struggle to infer runtime behaviors and expected outcomes. Regarding test effectiveness, we observe that the average mutation score across different LLMs consistently falls below 50\%, indicating that less than half of the introduced mutants are detected. This suggests that LLM-generated tests still have significant room for improvement in enhancing fault detection capability. 

\header{Application} The tests generated by \toolname operate under the assumption that they are created within a regression testing setting, where the correctness of the underlying focal method implementation is presumed. As \toolname supports both augmenting existing test classes and generating new ones, it can be integrated with any test generation tool to further improve code coverage in their generated test suites. Our evaluation focuses on Java SE projects from the Defects4J benchmark, which are compatible with Java 8 and newer versions. We select this benchmark because it includes projects with rich control-flow complexity, aligning with our goal of improving coverage for hard-to-cover branches. Consequently, \toolname is most applicable to Java SE projects with similar structural complexity. Extending the approach to Java EE applications, which pose additional challenges such as dependency injection and container management, remains an important direction for future work.

\header{Runtime} In our evaluation of 2,971 methods across 130 classes, the average runtime for \toolname to generate a passing test suite is 2.3 minutes per method, or approximately 53 minutes per class. Runtime varies with class complexity, method count, and project characteristics. For example, in Cli-40f, classes contain an average of 15 methods with a maximum cyclomatic complexity of 11, resulting in 0.8 minutes per method (12 minutes per class). In contrast, in Collections-28f, classes average 44 methods with cyclomatic complexity up to 32, leading to 9.4 minutes per method (111 minutes per class). Large utility classes with extensive iteration and map usage typically lead to longer generation times. The longer runtimes reflect the challenging nature of the selected classes, which include hard-to-cover branches and high structural complexity. Improving runtime efficiency remains an important avenue for future work.

	\section{Threats To Validity}

\header{Internal Validity} The effectiveness of \toolname depends on the capabilities of the LLM used, and results may vary if more advanced models are introduced in the future. To mitigate this, we evaluate \toolname using four recently released LLMs, including two open-source and two proprietary models. Another potential issue is the relationship between code coverage and test effectiveness, as high coverage does not necessarily imply meaningful assertions. To address this, we include mutation score in our evaluation to assess fault-detection capability alongside traditional coverage metrics. Additionally, \toolname integrates Comex~\cite{comex:arxiv2023} to generate control flow graphs for path extraction. Any bugs or limitations in Comex could potentially influence our results. To mitigate this risk, we verify the extracted paths during the implementation phase to ensure their correctness before they are used for test generation. Another concern is data contamination, i.e., the possibility that LLMs may have encountered parts of the Defects4J codebase during pretraining. We recognize this as a broader challenge in LLM-based software engineering research. To mitigate its effect and ensure fair comparisons, all evaluated prompting strategies use the same set of projects and the same underlying LLM. Our evaluation therefore reports relative improvements between prompting strategies rather than absolute performance.

\header{External Validity} Our current implementation and evaluation focus on Java, which may impact the generalizability of our findings. However, both LLM-based test generation and program analysis techniques used in \toolname are language-agnostic. The integration of dynamic code coverage analysis and static control flow analysis can be extended to support multiple programming languages, provided that equivalent tooling is available.

\header{Reproducibility} To ensure reproducibility, we have made \toolname's implementation and dataset publicly available~\cite{repo}. Additionally, we provide detailed instructions for replicating our experimental results.
	\section{Related Work}
Automated unit test generation has been a longstanding research topic, with approaches historically falling into two main categories: search-based and learning-based. Popular search-based approaches, such as EvoSuite \cite{evosuite:fse11, fraser2013evosuite} for Java and Pynguin \cite{pynguin, lukasczyk2020automated} for Python, utilize evolutionary algorithms to generate high-coverage unit tests. However, these approaches often suffer from generating test cases that are difficult for developers to understand and may lack meaningful assertions \cite{rao2023cat, utgen}. On the other hand, learning-based approaches leverage deep learning techniques to improve test case quality. For instance, ATLAS \cite{atlas} employs a neural machine translation model to automatically generate meaningful assertion statements for test methods, while \textsc{AthenaTest} \cite{tufano2020unit} uses a transformer-based model trained on developer-written test cases. These approaches produce more readable and accurate test cases, addressing key limitations of search-based methods.

With the rapid evolution of Large Language Models (LLMs), their ability to generate code has significantly improved, leading to their widespread adoption for automated test generation related tasks. Meta's TestGen-LLM~\cite{alshahwan2024automated} employs LLMs to automatically enhance existing human-written tests. \textsc{Libro}~\cite{libro} generates test cases directly from bug reports. Cedar~\cite{cedar:icse23} introduces an automated retrieval-based demonstration selection method to improve test assertion generation. CodaMosa\cite{lemieux:codamosa:icse23} enhances the search-based test generation tool Pynguin\cite{pynguin, lukasczyk2020automated} by incorporating LLMs to overcome coverage plateaus. UTGen~\cite{utgen} integrates search-based software testing with LLMs to improve the understandability of unit tests generated by EvoSuite. ChatUniTest~\cite{chen2024chatunitest} follows a generation-validation-repair mechanism to refine LLM-generated tests. ASTER~\cite{pan2025asternaturalmultilanguageunit} describes a generic pipeline applicable to multiple programming languages and complex software systems that require environment mocking. Empirical studies have also been conducted to evaluate the test generation capabilities of different LLMs across various programming languages, including JavaScript~\cite{schafer2023empirical}, Java~\cite{yang2024evaluation, siddiq2024using}, and Python~\cite{wang2024testeval}.

However, the empirical studies~\cite{siddiq2024using, schafer2023empirical, elhaji:test-generation-using-copilot:thesis23, wang2024testeval} highlight a persistent challenge: LLMs struggle to generate unit tests for complex methods and often fail to achieve sufficient branch coverage, particularly in real-world projects. To address this limitation, recent research has explored strategies to provide LLMs with additional program-related information beyond just the focal class or focal method, enhancing their ability to generate tests for complex code. 

HITS \cite{hits} prompts LLMs to decompose focal methods into slices and generate tests for each slice. However, these "slices" are not derived from traditional program slicing techniques \cite{weiser1984program} but instead represent arbitrary method fragments generated by LLMs. 

SymPrompt~\cite{symprompt} employs symbolic execution to extract path constraints from each method under test. It then reduces the extensive set of extracted paths to a manageable subset of constraints and generates tests incrementally, path by path. In contrast, \toolname leverages a path approximation algorithm that directly computes a linearly independent set of execution paths from the control flow graph. Moreover, while SymPrompt provides only branch conditions, \toolname delivers full execution paths, enabling the LLM to more accurately infer the correct test inputs.


Additionally, both SymPrompt and HITS provide static prompts containing method fragments (slices or paths) but lack guidance on where additional tests are needed, restricting them to generating new test classes rather than augmenting existing ones. CoverUp~\cite{pizzorno2024coverup}, on the other hand, provides code coverage information to indicate which parts of the code remain uncovered but does not guide the LLM on how to cover those uncovered segments effectively.

Unlike existing techniques, \toolname adopts an iterative workflow that leverages hybrid program analysis to guide the LLM. It integrates dynamic code coverage analysis and static control flow analysis to systematically extract and rank uncovered paths. These paths then help the LLM in reasoning about where and how to generate tests. Also, unlike previous methods, our approach effectively supports both augmenting existing tests and generating new test classes when no tests exist for a class/method.



	\section{Conclusion}
This paper introduced \toolname, a novel technique for automated unit test generation that seeks to emulate how human developers create tests. It addresses the challenge of achieving high branch coverage in complex code, where large language models often fall short. \toolname employs an iterative approach, combining static control flow analysis to identify execution paths with dynamic code coverage analysis to pinpoint under-tested areas. This hybrid program analysis guides an LLM to generate targeted test cases, which are then validated and repaired, using feedback to refine subsequent test generation efforts. Our evaluations on complex classes from the Defects4J~\cite{just2014defects4j} benchmark show \toolname significantly outperforms existing methods in line and branch coverage, achieving 26\% higher line coverage and 23\% higher branch coverage compared to the state-of-the-art, demonstrating the effectiveness of its iterative, program analysis-guided strategy for improving test thoroughness.


Looking ahead, we plan to extend \toolname to support multiple programming languages, broadening its applicability across diverse software projects. Additionally, we aim to incorporate cross-class context to further reduce hallucination and improve the accuracy of generated test cases.

	\bibliographystyle{ACM-Reference-Format}
	\interlinepenalty=10000
	\bibliography{main}
	
\end{document}